\documentclass[sigconf]{acmart}

\usepackage[T1]{fontenc}
\usepackage[utf8]{inputenc}
\usepackage[all]{nowidow}
\usepackage{balance}
\usepackage{microtype}
\usepackage[scaled=.8]{beramono}
\usepackage{url}
\usepackage[abbreviations]{foreign}
\usepackage[colorinlistoftodos]{todonotes}
\usepackage{hyperref}
\usepackage[inline]{enumitem}
\usepackage{xspace}
\usepackage{listings}
\usepackage{natbib}
\usepackage{pifont} 
\usepackage{caption}
\usepackage{subcaption}
\usepackage[poorman]{cleveref} 

\copyrightyear{2022}
\acmYear{2022}
\setcopyright{none}
\acmConference[ICSE-NIER'22]{New Ideas and Emerging Results }{May 21--29, 2022}{Pittsburgh, PA, USA}
\acmBooktitle{New Ideas and Emerging Results (ICSE-NIER'22), May 21--29, 2022, Pittsburgh, PA, USA}
\acmPrice{15.00}
\acmDOI{10.1145/3510455.3512783}
\acmISBN{978-1-4503-9224-2/22/05}

\newcommand*{\bb}{\textsc{BreakBot}\@\xspace}

\newcommand*{\dmodel}{$\Delta$-model\@\xspace}

\newcommand*{\mrc}{\textsc{Maracas}\@\xspace}
\newcommand*{\japi}{\texttt{japicmp}\@\xspace}

\makeatletter
\let\orig@lstnumber=\thelstnumber

\newcommand\lstsetnumber[1]{\gdef\thelstnumber{#1}}
\newcommand\lstresetnumber{\global\let\thelstnumber=\orig@lstnumber}
\makeatother

\begin{document}

\lstdefinelanguage{diff}{
    basicstyle=\ttfamily\small,
    morecomment=[f][\color{gray}]{@@},
    morecomment=[f][\color{green}]{+\ },
    morecomment=[f][\color{red}]{-\ },
	breaklines=true,
}

\lstdefinelanguage{log}{
    basicstyle=\ttfamily\small,
    breaklines=true,
	rulecolor=\color{black},
}

\lstset{
    language=Java,
	frame=tb,
	escapechar=@,
}

\title[\bb: Analyzing the Impact of Breaking Changes to Assist Library Evolution]{\bb: Analyzing the Impact of Breaking Changes\\to Assist Library Evolution}

\author{Lina Ochoa}
\email{l.m.ochoa.venegas@tue.nl}
\affiliation{
	\institution{Eindhoven University of Technology}
	\city{Eindhoven}
	\country{Netherlands}
}
\author{Thomas Degueule}
\email{thomas.degueule@labri.fr}
\affiliation{
	\institution{Univ. Bordeaux, Bordeaux INP, CNRS, LaBRI, UMR5800}
	\city{Talence}
	\country{France}
}
\author{Jean-Rémy Falleri}
\email{jean-remy.falleri@labri.fr}
\affiliation{
	\institution{Univ. Bordeaux, Bordeaux INP, CNRS, LaBRI, UMR5800, Institut Universitaire de France}
	\city{Talence}
	\country{France}
}

\begin{abstract}
``\textit{If we make this change to our code, how will it impact our clients?}''
It is difficult for library maintainers to answer this simple---yet essential!---question when evolving their libraries.
Library maintainers are constantly balancing between two opposing positions:~make changes at the risk of breaking some of their clients, or avoid changes and maintain compatibility at the cost of immobility and growing technical debt.
We argue that the lack of objective usage data and tool support leaves maintainers with their own subjective perception of their community to make these decisions.

We introduce \bb, a bot that analyses the pull requests of Java libraries on GitHub to identify the breaking changes they introduce and their impact on client projects.
Through static analysis of libraries and clients, it extracts and summarizes objective data that enrich the code review process by providing maintainers with the appropriate information to decide whether---and how---changes should be accepted, directly in the pull requests.
\end{abstract}

%
%

\maketitle

\section{Introduction}

Leveraging the time-honored principles of modularity and reuse, modern software systems development typically entails the use of external \emph{software libraries}.
Rather than implementing new systems from scratch, developers incorporate libraries that provide functionalities of interest into their projects.
Libraries expose their features through \emph{Application Programming Interfaces} (APIs) which govern the interactions between client projects and libraries.

Libraries constantly evolve to incorporate new features, bug fixes, security patches, refactorings, and extra-functional improvements~\cite{bavota2013evolution,brito2018and}.
It is critical for clients to stay up-to-date with the libraries they use to benefit from these improvements and to avoid security issues, technical lag, and technical debt~\cite{chinthanet2021lags,gonzalez2017technical}.
When a library evolves, however, it may break the contract previously established with its clients by introducing \emph{Breaking Changes} (BCs) in its public API, resulting in compilation-time, link-time, or run-time errors.
Seemingly innocuous changes such as altering the visibility or abstractness modifier of a type declaration or inserting a new abstract method can, under certain conditions, break client code. 
Errors triggered by these changes burden client developers, given the sudden urgency to fix issues out of their control without intrinsic motivation.
As a result, clients sometimes hesitate to upgrade their dependencies, raising security concerns and making future upgrades even more difficult.
Thus, it does not come as a surprise that the problem of helping clients respond to library evolution has garnered considerable interest in recent years~\cite{xing2007api,10.1145/1932682.1869486,6062100,6227179,lawall2018coccinelle}.

Surprisingly, however, the problem of helping library maintainers anticipate the impact of their changes and plan accordingly---the other side of the coin---has received little attention.
Libraries are generally accountable to their clients for providing stability~\cite{bogart2021when}.
We claim that library maintainers currently lack the necessary information and tool support to live up to this responsibility.
Indeed, library maintainers have limited means to foresee the consequences of their actions on their clients~\cite{hora2018developers,robbes2012developers} and would benefit from knowing precisely how their APIs are used in client projects.

The consequences of breaking an API declaration used by many clients, by a popular library, or by commercial clients may be dramatic:~clients may lose trust and migrate to another library.
As a result, some very cautious maintainers refrain from changing existing declarations, thinking it might be too impactful, leading to an accumulation of technical debt and an aging design that drive developers away~\cite{bogart2021when}.
Some more adventurous maintainers will push any change, disregarding all clients using the affected declaration, at the risk of breaking too many clients and having to revert the change ultimately~\cite{7816476}.
We believe that the sweet spot lies at the crossroads of innovation and stability.
Unfortunately, library maintainers have limited means to acquire evidence-based knowledge about the use of their API, and they often rely on their own judgment and subjective knowledge of their community to figure out how to evolve it~\cite{zhang2020enabling}.

In order to assist maintainers in the evolution of their libraries while avoiding breaking their clients unexpectedly, we propose the following vision: 
\emph{empowering library maintainers to make evidence-backed decisions regarding the evolution of their libraries by accurately and efficiently informing them about breaking changes and their impact on relevant clients.}
Concretely, this will enable maintainers to ask \emph{what-if} questions and explore what consequences evolving their library would have on their clients:~``\textit{What if we push this refactoring?}'', ``\textit{What if we alter this method's signature?}'', \etc.

In this paper, we introduce a proof-of-concept implementation of our vision: \bb, a GitHub bot that analyzes the content of Pull Requests (PRs) in Java repositories and employs static analysis to identify BCs and their impact on relevant clients.
The resulting reports are fed back into the PRs, enabling maintainers to quickly review changes and their impact and decide whether they should be accepted.
Because our tool employs static analysis, we focus on source- and binary-incompatible changes, which are an important class of BCs as they force clients to modify their code at upgrade time, and leave behavioral changes aside.
We expect our vision and its implementation in \bb to fight immobility and stagnation by pushing maintainers to introduce BCs when it is safe and avoid pushing these changes when they are too impactful.
Making the co-evolution of library and clients more harmonious should benefit library maintainers and client developers alike.

\section{Background \& Motivating Example}
\label{sec:motivation}

Deciding whether and when to introduce BCs is no trifling matter. 
The decision depends on values and constraints established at the ecosystem and library levels. 
However, the desire to preserve \textit{stability} and to prioritize the sense of \textit{community} are generalized concerns among all projects in most ecosystems~\cite{bogart2021when}. 
To enact stability, libraries tend to assume that BCs are always harmful~\cite{javan2021dependency}. 
This reasoning  has led to the misconception that stability is equivalent to change stagnation~\cite{bogart2021when, businge2019stable, mcdonnell2013empirical}.
However, we claim that stability is the capability to not break clients even under the presence of change.
What truly affects stability are \emph{changes that break clients}, not BCs themselves.
This distinction is essential:~it has been shown that, for instance in Maven Central, most BCs do not impact the clients whatsoever~\cite{ochoa2021breaking}.
Another problem strikes when evidence about how changes are impacting clients is not communicated, increasing the mistrust in the community. 
Thus, some evolving libraries opt for communicating change via their API, versioning conventions such as semantic versioning, or official documentation such as release notes and changelogs.
However, these artifacts might be seen as promises that, when broken, result in further mistrust.
Furthermore, official documentation tends to be incomplete given that only notable changes are requested to be included~\cite{keepachangelog}.

To illustrate the issues that library maintainers run into when introducing BCs, we take an example brought to our attention by a core maintainer of Spoon---a source code analysis and transformation library for Java developed on GitHub~\cite{spoonspe}.
Spoon developers are very concerned about stability and therefore avoid BCs.
When they do want to introduce a BC, they use deprecation to warn clients before proceeding with the change.
In pull request \href{https://github.com/INRIA/spoon/pull/2683}{PR\#2683} a new feature to pretty-print Java import statements was rolled in, making a previous implementation---the \texttt{ImportScanner} interface and corresponding \texttt{ImportScannerImpl} implementation---obsolete.
Therefore, the author of the PR labeled both types as deprecated. 
Two months later, in \href{https://github.com/INRIA/spoon/pull/3184}{PR\#3184}, Spoon developers removed a set of deprecated types and methods from the code, notably the two types mentioned above.
The PR was merged, making its way into Spoon's main branch.
However, three months later, Spoon's developers noticed that these changes broke two important clients (Astor~\cite{martinez2016astor} and DSpot~\cite{danglot2019dspot}) and realized that the \texttt{ImportScanner} interface and implementation were in fact used by clients and thus reconsidered the deprecation.
They finally reintroduced them in \href{https://github.com/INRIA/spoon/pull/3266}{PR\#3266}.

It is evident that having to revert changes after introducing them is not ideal.
If other changes had been made based on the original changes, they would have had to be reverted too, triggering a displeasing ripple effect within the library's code.
How could this have been avoided?
To the best of our knowledge, there are three approaches enabling library maintainers to evaluate the BCs they introduce.

\paragraph{Regression testing}
The traditional approach is to use a regression test suite.
However, a regression test suite cannot cover every possible use of the library, and therefore unanticipated BCs may leak to the clients.
In addition, when the test suite detects a regression, it is not able to evaluate its impact on the library's clients.
Maintainers must resort to their subjective opinion to decide whether or not to include the change.
In our example case, Spoon's test suite noticed the BCs introduced in PR\#3184, and the \texttt{ImportScannerTest} class was deleted as part of the same PR.
However, since the deleted types were deprecated, the maintainers did not reconsider the change.

\paragraph{Static analysis}
Various tools can scan two versions of a library using static analysis to output the list of BCs between them~\cite{jezek2017api}.
This approach is more complete than regression testing (although false positives or negatives can arise), therefore it is unlikely that BCs will leak to clients.
For instance, Guava relies on JDiff~\cite{guavajdiff} while the Apache foundation uses \japi~\cite{apachejapicmp} to systematically search for BCs.
Nevertheless, these tools only list the introduced BCs without showing the actual impact on client projects.
If Spoon's developers were using this type of tool, it would have noticed the changes the same way their test suite did.

\paragraph{Reverse dependency compatibility testing}
The third approach---the only one able to evaluate the impact on clients---is Reverse Dependency Compatibility Testing (RDCT)~\cite{DBLP:phd/hal/Zimmermann19}.
The idea of RDCT is to identify a set of relevant clients, retrieve their source code, inject the new version of the library in their dependencies, and finally build them and run their test suite.
If any error is detected during the process, maintainers are notified and can analyze the issue by browsing through the build log files.
Projects like Scala, Rust, or Spoon perform periodical RDCT on their clients.
The Coq project goes a step further by directly integrating RDCT as part of its Continuous Integration (CI) process, so that any BC impacting a client is detected right away~\cite{zimmermann2021extending}.

However, RDCT suffers from several drawbacks.
First, selected clients must always be buildable and have a test suite that passes on the previous version of the library.
Second, RDCT takes a considerable amount of time and computational resources to build and test every client.
For instance, the Rust project reports requiring up to a week to check 74,234 clients, as of September 2019.
Third, when BCs are introduced, the resulting log files are verbose, forcing the maintainers to filter irrelevant information to understand the root cause.
Last, whenever a BC is merged, the impacted clients can no longer be used for subsequent RDCT until they are repaired.
For this reason, Coq's maintainers resort to fixing the clients themselves to keep their CI process running, which requires considerable effort.

\begin{lstlisting}[language=log, numbers=left, xleftmargin=4em, framexleftmargin=4em, caption=Extract of Astor's build log file (704 lines in total)., label=lst:astor, basicstyle=\ttfamily\footnotesize, float=t, belowskip=-14pt]
Started by timer@\lstsetnumber{\ldots}@
@\lstresetnumber\setcounter{lstnumber}{210}@
[ERROR] The goal you specified requires a project to execute but there is no POM in this directory. @\lstsetnumber{\ldots}@
@\lstresetnumber\setcounter{lstnumber}{225}@
[ERROR] Entities.java:[63,29] unmappable character for UTF-8 @\lstsetnumber{\ldots}@
@\lstresetnumber\setcounter{lstnumber}{294}@
[ERROR] TypeUtilsTest.java:[524,47] incompatible types: inferred type does not conform to upper bound(s)@\lstsetnumber{\ldots}@
@\lstresetnumber\setcounter{lstnumber}{653}@
@\bfseries [ERROR] LibParser.java:[23,29] cannot find symbol@
@\bfseries [ERROR] \quad symbol:   class ImportScanner@
\end{lstlisting}

In our example case, Spoon's RDCT runs every night on 13~projects.
Following PR\#3184, it ran into errors for the Astor and DSpot clients.
An analysis of the log file of Astor (\Cref{lst:astor}), containing 266 errors among 704 log statements, revealed that the deleted types raised missing symbol errors at compile-time.
Thanks to RDCT, Spoon developers discovered the impact of the BC.
However, Astor's RDCT build was failing for three months before the revert and thus could not be used to check for other BCs in this period.

In summary, while RDCT helps analyze the impact of BCs on clients, its drawbacks severely hinder its usability and adoption.
In the next section, we present a novel and lightweight approach based on static analysis to evaluate the impact of source- and binary-incompatible BCs on clients and integrate it within the maintainers' workflow.
Naturally, RDCT still goes beyond our approach regarding behavioral-incompatible changes, at the expense of building and running the client test suites.

\section{\bb}
\label{sec:bb}
\bb (\url{https://github.com/alien-tools/breakbot}) is a proof-of-concept implementation of our vision.
It analyzes libraries and clients written in Java and integrates with GitHub's pull-based development workflow.
It is implemented as a GitHub bot that provides library maintainers with the right information whenever and wherever needed.
Making \bb a bot allows us to fully automate the process and be as close as possible to the maintainers' development workflow.
Because \bb feeds helpful information directly in GitHub's PRs, we expect it to enhance the code review process, as was shown for other bots in previous work~\cite{wessel2020effects}.

\subsection{Overview}
\begin{figure}[tb]
	\includegraphics[width=.85\linewidth]{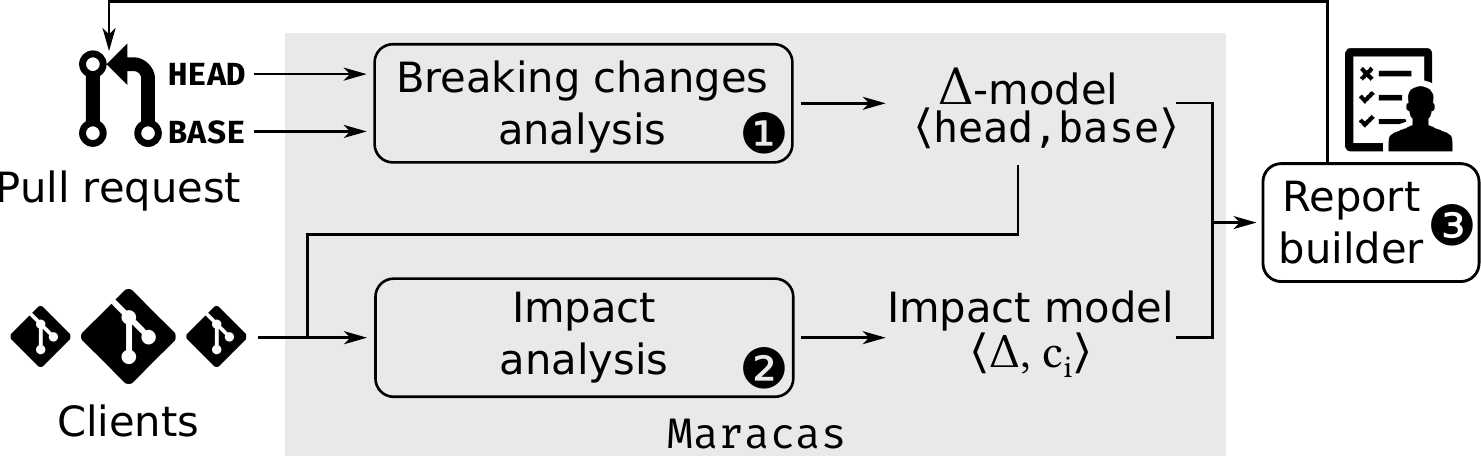}
	\caption{\bb handles the communication between GitHub and \mrc to produce meaningful reports.}
	\label{fig:overview}
\end{figure}

Once maintainers install \bb on a library's repository and configure a list of clients of interest, \bb registers itself via the GitHub's Checks API~\cite{githubchecks} to analyze each new or updated PR.
Then, it reports:~i)~the list of BCs they introduce, and ii)~the impact of those BCs on the configured list of clients.
Under the hood, \bb leverages a novel static analysis tool, \mrc~\cite{maracas}, to analyze libraries and clients and to infer the list and impact of BCs~\cite{ochoa2021breaking}.
Note that \bb's design enables it to be reused with little effort for other analysis frameworks similar to \mrc, targeting other programming languages.
\Cref{fig:overview} gives a high-level view of \bb.
We detail below the main steps of the process.

\paragraph{\ding{202} Breaking changes analysis}
In a first step, \bb communicates to \mrc the GitHub PR to analyze.
In every PR, the \texttt{head} pointer points to the branch containing the PR's code, while the \texttt{base} pointer points to the branch where the changes should be pulled into.
The diff between \texttt{head} and \texttt{base} represents the changes introduced by the PR.
\mrc first clones and builds both these versions to obtain binary JARs, uses \japi to extract the list of BCs, and stores it in a dedicated \dmodel~\cite{ochoa2021breaking}.
Each BC in a \dmodel specifies the affected declaration (a particular type, method, or field) and the kind of BC (\eg method removed, field now final, class less accessible).
A configuration file hosted on the library's repository allows maintainers to specify how to build their library and which parts of their API should be excluded from the analysis because they are exempt from compatibility guarantees (\eg, \texttt{@Experimental} declarations and \texttt{*test*} packages).

\paragraph{\ding{203} Impact analysis}
In a second step, for each configured client, \mrc builds an impact model that pinpoints how the changes in the \dmodel impact the project.
While we originally designed \mrc to analyze the impact of BCs on bytecode~\cite{ochoa2021breaking}, we redesign it to statically analyze their impact on source code using the Spoon framework.
Because this new implementation does not require clients to be built, it frees \bb from the limitations of state-of-the-practice RDCT regarding build times and the need to repair clients constantly.
Each location in client code impacted by a BC is associated with:~i)~the impacted AST node in client code (\eg a variable, a method invocation, an import); ii)~the library declaration it uses that is impacted by a BC, and; iii)~the way the client is using the declaration (\eg through class instantiation or extension, as a type argument, through field access).

\paragraph{\ding{204} Report builder}
In a third step, the \dmodel and impact models are sent back to \bb, which builds a final report incorporated directly within the PR.
This report displays the result of the analysis in a way that library maintainers can easily interpret.
We design the report to highlight the essential information first:~the list of BCs, how each of them individually impacts client code, and how many of the clients are impacted.
Then, a more detailed report describes, for each client, every location in its source code impacted by the changes.
To make it easier to understand the causes and consequences of BCs, every code reference in the report directly links to the corresponding source code on GitHub.

\begin{figure*}[t!]
	\centering
	\parbox{\textwidth}{
		\parbox{.725\textwidth}{%
			\subcaptionbox{Excerpt of the breaking changes and their impact on clients\label{fig:report:bcs}}{\includegraphics[width=\hsize]{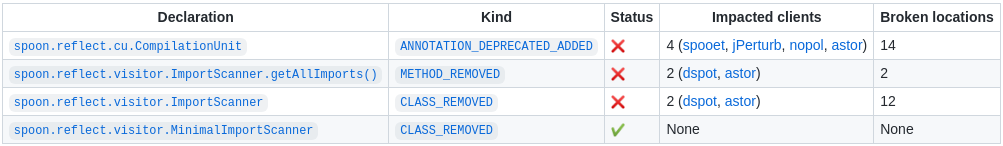}}
			\subcaptionbox{Excerpt of the impacted locations in Astor\label{fig:report:astor}}{\includegraphics[width=\hsize]{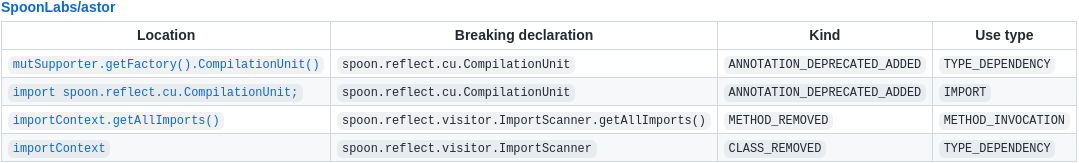}}  
		}
		\hskip.5em
		\parbox{.251\textwidth}{%
			\subcaptionbox{Excerpt of clients' overview\label{fig:report:clients}}{\includegraphics[width=\hsize]{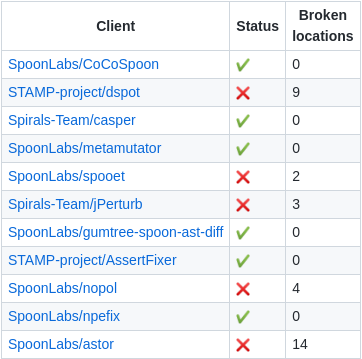}}			
		}
	}
	\caption{Excerpts of the report generated by \bb for Spoon's PR\#3184 on GitHub}
	\label{fig:report}
\end{figure*}

\subsection{Revisiting the Spoon Example}
We used \bb and \mrc to analyze PR\#3184 and showcase how it alleviates the drawbacks of RDCT identified in \Cref{sec:motivation}.
\Cref{fig:report} depicts the most important parts of the report.

First, because \mrc employs static analysis to infer the impact of BCs, it only needs to parse the clients to build their ASTs.
It does not require clients to be built and thus avoids the need for a complex CI system that may suffer from additional failures:~dependency resolution, flaky tests, \etc.
Besides, while Spoon's RDCT took 6~minutes to analyze Astor only, \bb could analyze all 13~clients in 2~minutes.

Second, RDCT suffers from the need to analyze verbose and cryptic log files for each client, separated from the PR's code, riddled with unrelated outputs and errors, and messages that do not explicitly map client errors to their root cause at the library level (\Cref{lst:astor}).
In contrast, the report built by \bb centralizes and highlights the essential information:~the PR introduces 13 BCs, causing 32 broken locations in client code, and 5 out of 13 clients are impacted by the changes (\Cref{fig:report:clients}).
The remainder of the report details, for each client, which particular locations are impacted to let the maintainers understand \emph{why} the clients break and how they might improve the PR (\Cref{fig:report:astor} for Astor).

In this particular case, \bb accurately reports that the class removals impact Astor and DSpot, but also that the type \texttt{CompilationUnit}, deprecated in the same PR, is used by clients (\Cref{fig:report:bcs}).
This helps maintainers reflect on whether deprecating this type is appropriate to avoid making the same mistake again.
The removal of the \texttt{MinimalImportScanner} type, however, does not impact any client and is thus innocuous.
The final report effectively answers the \emph{what-if} questions stemming from the PR.



\section{Future Plans}
\label{sec:future}

Our first implementation of the vision, \bb, is a mere stepping stone.
We intend to implement the following research roadmap.

\paragraph{Configurability.} 
Policies to deal with backward compatibility range from permissive (\eg npm, favoring \textit{innovation}) to restrictive (\eg Eclipse, favoring \textit{compatibility})~\cite{bogart2021when}. 
Regardless of their posture, projects should be given the necessary information to decide whether to introduce a change. 
Maintainers, therefore, need highly configurable impact analysis tools to anticipate the impact of their changes.
\textit{What are}, then, \textit{the configuration properties that impact analysis tools should capture?}
For instance, defining the type and number of allowed BCs, the parts of the API subject to analysis (\eg non-experimental interfaces), and the key set of relevant clients (\eg commercially-related projects), is of foremost importance.

\paragraph{Discovery of clients.} 
Knowing how BCs impact relevant clients is key when deciding to apply or revert changes.
Some projects, such as Coq and Spoon, have a precise list of relevant clients, but not all projects do.
Even when such a list exists, it might be incomplete and hide important evidence. 
Automatically discovering a \emph{diverse and representative sample}~\cite{nagappan2013diversity} of clients should give a clearer picture of the library's usage.
To do so, one must first \textit{pinpoint which versions of the library are affected by a given BC}, and thus \textit{which clients may be affected}.
As it is likely that clients' usage of a library will overlap~\cite{harrand2022api}, it is essential to \textit{avoid analyzing similar clients multiple times}, \eg by clustering clients around their usage of the library.

\paragraph{Reporting actionable insights.} 
\bb identifies the list of library declarations introducing BCs, and the list of impacted locations in client code.
Although it improves the state of practice, it may still be hard for developers to interpret this information.
Regarding BCs, \textit{should they be represented as atomic modifications to the code, or should more meaningful operations that show the developer's intentions (\eg refactorings) be listed instead?}
Then, \textit{what is the right level of granularity to represent impacted client code (\eg at the declaration, line, or AST node level)?}
Finally, to decide whether to introduce a BC, \textit{how to measure its criticality?}
\textit{How to define proper thresholds for metrics to enact project-specific policies?}

\paragraph{Approach evaluation.}
To evaluate \bb, we plan to use both quantitative and qualitative evaluation methods.
We will first investigate the accuracy of \mrc during the BCs and impact analysis phases using a proper benchmark such as the one introduced by \citet{jezek2017api}.
Second, we aim at better understanding the current state of practice when dealing with change and the effects of \bb on these practices.
Therefore, we plan to deploy \bb on the GitHub repositories of a few libraries (\eg Spoon, ASM, JavaParser, Vallang, GumTree) and interview their maintainers.
In particular, we plan to inquire whether \bb improves the analysis of code changes during code reviews.
\textit{Does \bb ``scare'' developers more, or does it encourage them to make changes they were afraid of because they could not foresee their impact?}


\begin{acks}
We thank Théo Zimmerman, Matias Martinez, and Jurgen Vinju for their feedback on earlier versions of this manuscript, and Léonard Rizzo for the initial implementation of \bb.
This work was partially funded by the French National Research Agency through grant ANR ALIEN (ANR-21-CE25-0007).
\end{acks}

\clearpage
\balance
\bibliographystyle{ACM-Reference-Format}
\bibliography{main}


\end{document}